\documentclass{article}

\usepackage{cite}
\usepackage{graphicx}
\usepackage{caption}
\usepackage{amssymb}
\usepackage{array}
\usepackage{dcolumn}
\usepackage{bm}
\usepackage{textgreek}
\usepackage{amsmath}
\usepackage[letterpaper, total={7in, 9in}]{geometry}
\usepackage{authblk}

\newcommand*\dbar[1]{\overline{\overline{#1}}}

\title{Integrated optical isolators using electrically driven acoustic waves}

\author[1,*]{Nathan Dostart}
\author[1,2]{Yossef Ehrlichman}
\author[1,3]{Cale Gentry}
\author[4]{Milo\v s Popovi\'c}

\affil[1]{Dept. of Electrical, Computer, and Energy Engineering, University of Colorado, Boulder, CO, 80309, USA}
\affil[2]{Currently with Axalume, San Diego, CA, USA}
\affil[3]{Currently with SRI International, 2595 Canyon Blvd.
Suite 440, Boulder, CO 80302, USA}
\affil[4]{Dept. of Electrical and Computer Engineering, Boston University, Boston, MA, 02215, USA}

\affil[*]{Corresponding author: nathan.dostart@colorado.edu}

\date{\today}

\begin{document}
	
\maketitle

\begin{abstract}
We propose and investigate the performance of integrated photonic isolators based on non-reciprocal mode conversion facilitated by unidirectional, traveling acoustic waves. A triply-guided waveguide system on-chip, comprising two optical modes and an electrically-driven acoustic mode, facilitates the non-reciprocal mode conversion and is combined with modal filters to create the isolator. The co-guided and co-traveling arrangement enables isolation with no additional optical loss, without magnetic-optic materials, and low power consumption. The approach is theoretically evaluated and simulations predict over 20 dB of isolation and 2.6 dB of insertion loss with 370 GHz optical bandwidth and a 1 cm device length. The isolator utilizes only 1 mW of electrical drive power, an improvement of 1--3 orders of magnitude over the state-of-the-art. The electronic driving and lack of magneto-optic materials suggest the potential for straightforward integration with the drive circuitry, possibly in monolithic CMOS technology, enabling a fully contained `black box' optical isolator with two optical ports and DC electrical power.
\end{abstract}


\section{Introduction}
\label{sec:intro}

Integrated photonics is rapidly advancing and will become an integral part of future computing and communication technologies, including through co-integration of state-of-the-art electronics and silicon photonics \cite{sun2015single}. In particular, monolithic co-integration of electronics and photonics has been demonstrated in advanced process nodes ($28-45\,$nm) \cite{chen201522,sun2015single} and combining this technology with on-chip lasers has been a driving goal of integrated photonics. Any effective use of on-chip lasers requires adjacent, on-chip isolators to protect the lasers from downstream back-reflections and enable high performance comparable to current (off-chip) commercial systems. Even for architectures that use off-chip lasers, on-chip isolators or circulators could enable simultaneous bi-directional communication links and other capabilities. Conventional isolators use the magneto-optic (Faraday) effect to induce non-reciprocity, but magneto-optic materials have substantial loss and have been challenging to integrate in photonic platforms \cite{ghosh2012yig,stadler2014integrated,huang2016electrically,huang2017integrated}. Despite these efforts, and concurrent work using other non-reciprocal effects, to our knowledge no silicon photonics foundry currently has an optical isolator in its component library. Hence, non-magnetic approaches to integrated photonic isolation, preferably CMOS-compatible, are desirable and would be rapidly adopted.

Aside from the magneto-optic effect, two other avenues exist to produce a non-reciprocal system: nonlinearity and time variance \cite{jalas2013and}. Optical nonlinearity uses effects such as saturable absorption/gain \cite{chang2014parity} or nonlinear optical coupling due to e.g. the Kerr effect \cite{fan2012all,ramezani2010unidirectional} or nonlinear thermal effects \cite{li2020reconfigurable}, but all such isolators are inherently nonlinear in the optical signal. This immediately precludes such isolators from use with modulated (signal-carrying) light, as well as issues with protecting the isolated component for a certain class of low-amplitude reflections \cite{shi2015limitations}.

Time-varying optical permittivity \cite{sounas2017non} avoids these issues but requires active modulation of the permittivity. One potential approach to create this time-variance is optoelectronics,  using effects such as frequency conversion \cite{doerr2014silicon,tzuang2014non,yang2014experimental}, traveling wave phase shifters \cite{ibrahim2004non,dong2015travelling}, and optical interband transitions \cite{yu2009complete,lira2012electrically}. The main drawback of current silicon photonic approaches is the free carrier loss inherent to plasma-dispersion modulators \cite{soref1987electrooptical}, with the only three optoelectronic isolators demonstrated in silicon photonic platforms so far having isolation ratios much lower than needed by applications ($2-5\,$dB \cite{lira2012electrically,doerr2014silicon,tzuang2014non}).

Acousto-optic modulation (Brillouin scattering), where an acoustic wave modulates the optical signal, can be used to create a time-varying optical permittivity without optical loss. There have been a couple such demonstrations so far in silicon \cite{fang2017generalized,kittlaus2018nonreciprocal} using a pump laser to induce stimulated Brillouin scattering (SBS). However, the use of a pump laser effectively negates any utility of these isolators to isolate a separate, on-chip laser.

In order to avoid the downsides of previous demonstrations (pump lasers, signal-dependent transmission, and low isolation) we propose to use transducers to drive an acoustic wave to generate non-reciprocity. On-chip transducers have been extensively demonstrated \cite{li2015nanophotonic,balram2016coherent,sohn2018time}, providing a low-risk method of exciting an acoustic wave without optical pumping and simultaneously avoiding the high losses of plasma-dispersion modulation. One recent demonstration \cite{sohn2018time} utilized transducers to create 15 dB of isolation in a resonator (1 GHz bandwidth) fabricated on AlN. Our proposed design provides an analogous, broadband isolator design where co-propagating optical and acoustic modes enable efficient acousto-optic modulation with low acoustic drive powers, small transducers, and the theoretical minimum device length. It benefits over this previous demonstration both in its bandwidth (100s of GHz) and power efficiency due to the co-propagating arrangement.

In this paper we investigate a class of integrated photonic isolators that utilize an optical guided wave `interband transition' i.e. spatial mode conversion \cite{yu2009complete} facilitated by non-reciprocal coupling induced by an electronically-excited, traveling, guided acoustic wave. We refer to them as electro-mechanical photonic (EMP) isolators. This approach avoids the requirement of an optical pump \cite{poulton2012design,kittlaus2018nonreciprocal,kang2011reconfigurable,kim2015non}, achieves lower insertion losses and higher isolation than carrier plasma-dispersion approaches \cite{lira2012electrically,tzuang2014non,doerr2014silicon}, and may use potentially co-integrated electronics to generate the acoustic wave to result in a self-contained device. Our design particularly benefits from a co-propagating, guided acoustic wave in a velocity-matched optical/acoustic waveguide to achieve a broad bandwidth and high energy efficiency.

\begin{figure}[t]
	\centering
	\includegraphics[width=\textwidth]{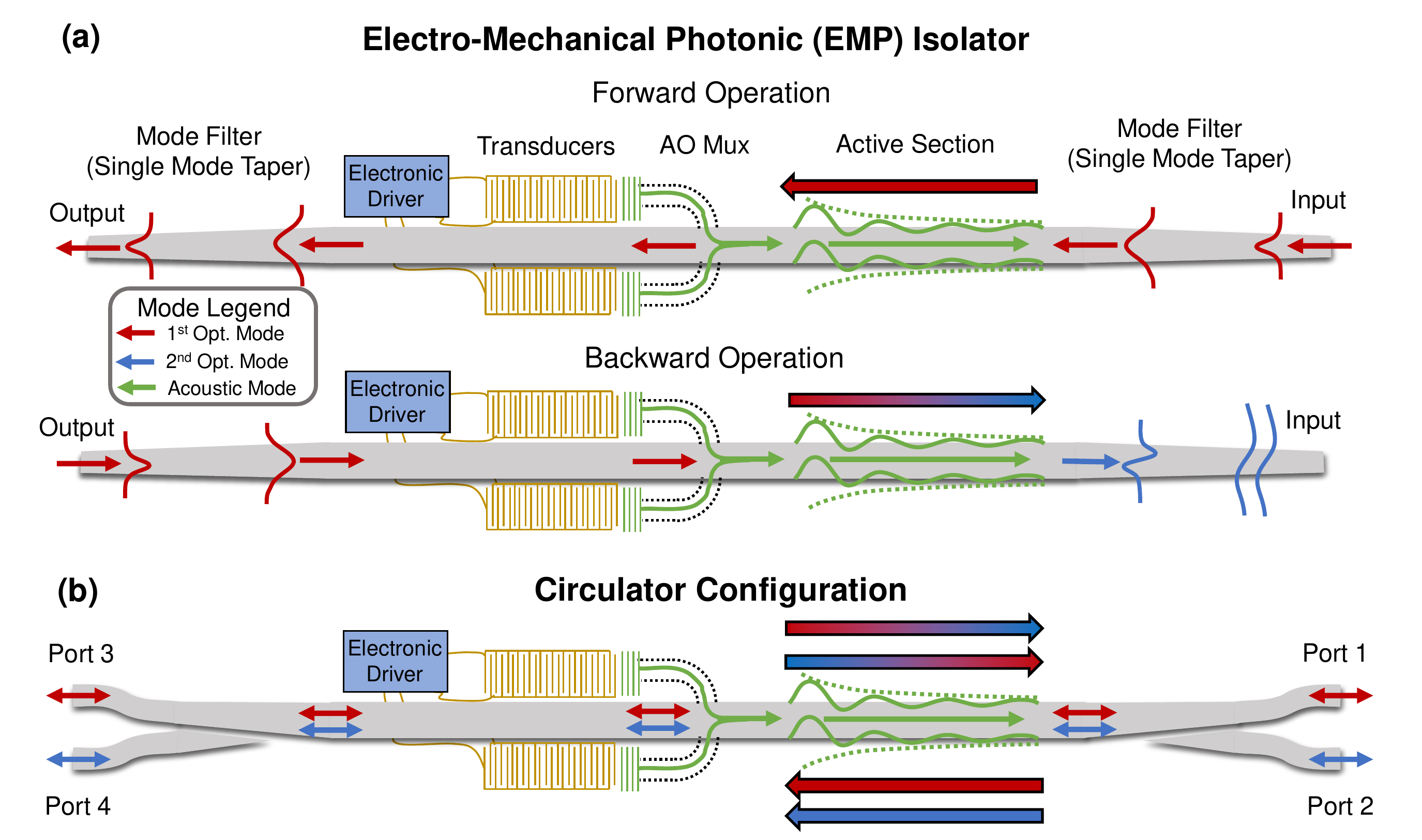}
	\caption{ (a) Schematic of the EMP isolator. Electrically-driven transducers generate an acoustic mode (green) which is injected into an `active' section, along with two optical modes (red and blue), by an acousto-optic multiplexer (AO mux) \cite{muxarxiv}. This active section facilitates optical mode conversion in the backward direction, but not in the forward direction, using the acoustic wave. On either side of the active section the second optical mode is filtered by the taper down to a single mode waveguide (the input and output ports). In the forward direction (top), the first optical mode propagates through the active section in the opposite direction as the acoustic wave and therefore is not mode-converted, exiting the isolator without significant power loss. In the backward direction (bottom), the first optical mode co-propagates with the acoustic wave and is entirely converted to the second mode, which is then filtered by the taper to provide isolation. (b) Circulator configuration, where the tapers are replaced by optical waveguide mode multiplexers/demultiplexers (see e.g. \cite{sun2016ultra}) for a 4-port circulator.}
	\label{fig:iso_concept}
\end{figure}

\section{EMP Isolator Concept}
\label{sec:concept}

The key components of the EMP isolator, shown schematically in Fig.~\ref{fig:iso_concept}(a), are a triply-guiding (two optical modes, one acoustic mode) waveguide cross-section (the `active' section) with strong optomechanical coupling and tailored dispersion, a transducer to drive the acoustic wave, and an acousto-optic multiplexer (AO mux) which combines the optical modes with the transduced acoustic mode.

The operating principle of the EMP isolator is simple: in the forward direction [Fig.~\ref{fig:iso_concept}(a,top)], the incident (first) optical mode enters through the input port, tapers out to the wider active section, traverses the active section unaffected, and is passed by the output taper. In the backward direction [Fig.~\ref{fig:iso_concept}(a,bottom)] the active section fully converts the first optical mode to the second optical mode, which is then filtered by the taper down to a single mode waveguide. The acoustic wave is injected into the active section without reflections by the AO mux \cite{muxarxiv} and terminated before reaching the taper either using the AO mux as a demux (filtering out the acoustic wave) or through decay due to coupling to other phonon modes \cite{rodriguez2019direct}. This non-reciprocal `interband' mode conversion in the active section, in combination with modal filtering, thereby provides isolation. When the tapers are replaced with optical multiplexers, the EMP isolator can act instead as a 4-port circulator [Fig.~\ref{fig:iso_concept}(b)] or a 3-port circulator if only one taper is replaced.

\begin{figure}[t]
	\centering
	\includegraphics[width=\textwidth]{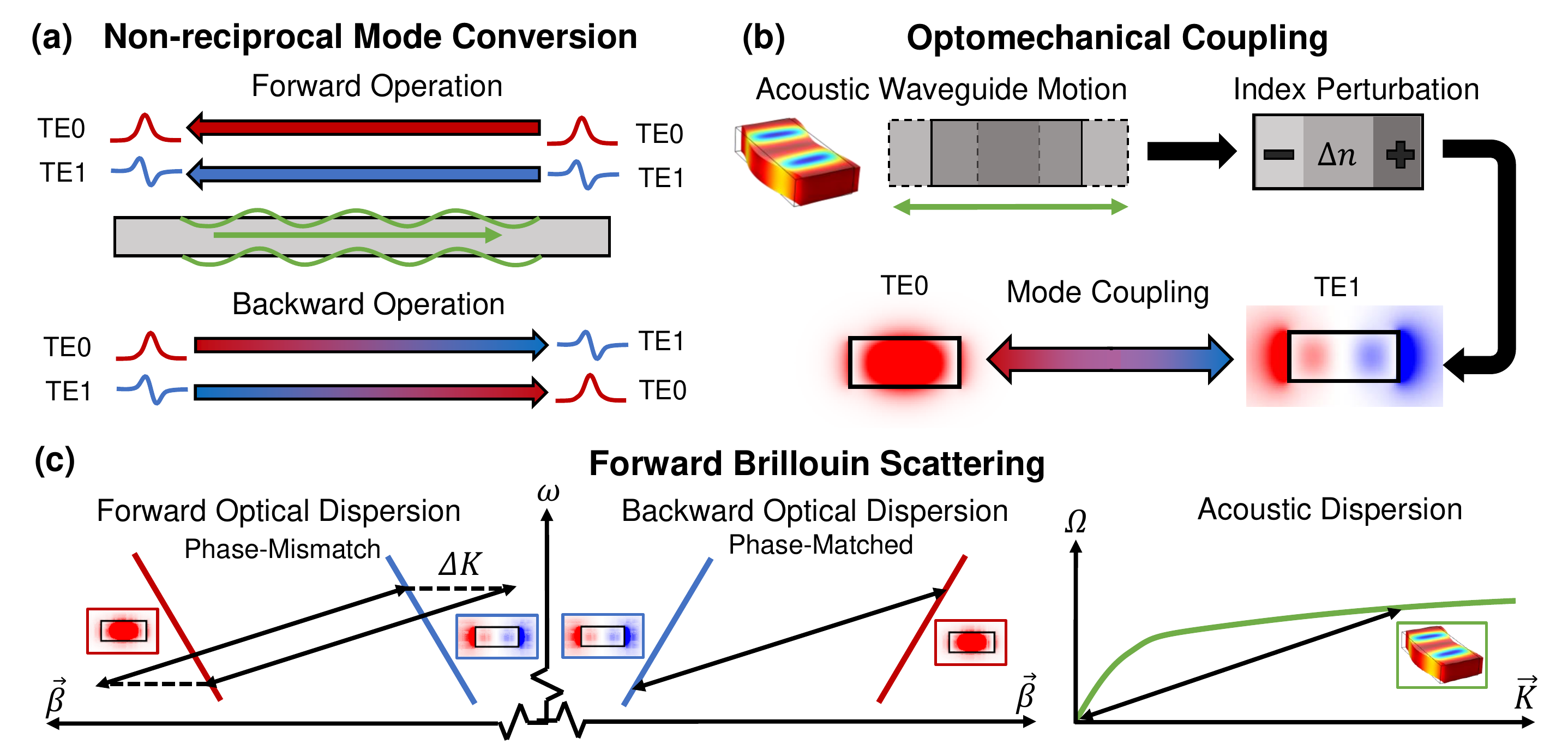}
	\caption{(a) The `active' section facilitates optical mode conversion in the backward direction, but not in the forward direction, using forward Brillouin scattering of the optical modes by the acoustic wave. (b) Optomechanical coupling induces optical mode conversion, where the acoustic wave displaces the waveguide to create a refractive index perturbation which couples the modes. (c) The unidirectionally propagating acoustic wave (right) is only phase-matched in the backward direction (center) and not in the forward direction (left), leading to non-reciprocal mode conversion.}
	\label{fig:nonrecip_conversion}
\end{figure}

The non-reciprocal mode conversion in the active section is provided by a linear optomechanical interaction. As shown in Fig.~\ref{fig:nonrecip_conversion}(a), the triply-guiding waveguide co-confines both light and sound such that the acoustic mode couples the incident optical mode and a second optical mode via Brillouin scattering \cite{wolff2015stimulated}. The active section converts all power from the first optical mode to the second optical mode (and vice versa) in the backward direction of the isolator; no mode conversion occurs in the forward direction. Optomechanical coupling, discussed in further detail in Sec.~\ref{subsec:om_coup}, is induced by the acoustic wave: acoustic displacement of the waveguide creates an index perturbation which couples the optical modes [Fig.~\ref{fig:nonrecip_conversion}(b)]. The non-reciprocal aspect of this coupling is due to the unidirectional propagation of the acoustic wave, which is phase-matched in the backward direction but not the forward direction [Fig.~\ref{fig:nonrecip_conversion}(c)]. For proper choice of active section length (see Sec.~\ref{subsec:device_length}), all light can be converted in the phase-matched direction but no light is converted in the phase-mismatched direction. This results in the desired non-reciprocal mode conversion by the active section.

The main innovation introduced by our approach is the use of co-guided optical and acoustic waves, whereas other approaches have used either non-propagating or transversely propagating acoustic waves. In both of the latter geometries, only a small portion of the acoustic power is utilized for acousto-optic modulation due to fast absorption or small overlap of the acoustic mode with the optical modes. The AO mux in particular enables this novel architecture as it unidirectionally injects the acoustic wave into the triply-guiding waveguide without affecting the optical modes \cite{muxarxiv}. Previous isolator designs, lacking such a multiplexer, were incapable of exciting a propagating acoustic mode guided in the same cross-section as optical modes. Such architectures \cite{sohn2018time,van2018electrical,li2015nanophotonic}, where the acoustic wave is not co-propagating and co-guided with the optical modes, can be used to create isolation in a similar manner to the co-propagating architecture proposed here but require longer transducers and more drive energy.

To achieve true co-propagation, we use a propagating phonon mode without a cutoff frequency. The typical phonons used in forward SBS interactions, by contrast, are generally chosen because of their phase velocity which can be made comparable to the speed of light. However, to achieve these high phase velocities they must be excited near their cutoff frequency at which their group velocity approaches 0 and the phonons effectively do not propagate. This condition corresponds to an exceptionally short absorption length, on the scale of the acoustic wavelength ($\sim10$ {\textmu}m). Use of these non-propagating phonons, with electrical excitation, therefore requires a distributed transducer as long as the active section, such as \cite{fan2016integrated}, and correspondingly high power consumption. Propagating phonons, on the other hand, have a much longer absorption length (on the order of a millimeter \cite{li2015nanophotonic}) allowing for lower acoustic powers and significantly more compact transducers.

Forward Brillouin scattering, where two co-propagating optical modes are coupled, is used here in order to make the mode conversion optically broadband. For the forward scattering configuration, phase-matching is fundamentally limited only by the relative dispersion of the two optical modes. In the backward scattering configuration, where two counter-propagating optical modes are coupled, the optical bandwidth is fundamentally limited to approximately the acoustic frequency (BW$\approx\Omega/2\pi$) because the two optical modes have group velocities with opposite signs.

\section{Analysis of the EMP Isolator}
\label{sec:math}

We use the coupled mode theory treatment of optomechanical interactions in \cite{wolff2015stimulated}. The optical and acoustic fields are represented in their respective modal bases, $\vec{E}=A_1(z)\vec{\mathcal{E}}_1(x,y)e^{j(\omega_1t-\beta_1z)}+A_2(z)\vec{\mathcal{E}}_2(x,y)e^{j(\omega_2t-\beta_2z)}$, $\vec{u}=B(z)\vec{\mathcal{U}}(x,y)e^{j(\Omega t-Kz)}$. Here $\vec{E}$ and $\vec{u}$ represent the total electric field and acoustic displacement field; $\vec{\mathcal{E}}_i$ and $\vec{\mathcal{U}}$ are the (vectorial) modal field shapes; $\omega_i$ and $\Omega$ are the mode frequencies; $\beta_i$ and $K$ are the modal propagation constants; and $A_i$ and $B$ denote the wave amplitudes (the modal shapes are normalized such that the powers are given by $|A_i|^2$, $|B|^2$).

\subsection{Coupled Mode Equations}
\label{subsec:coup_mode_eqs}

The coupled mode equations that describe interactions between these modes are \cite{wolff2015stimulated}

\begin{equation}
\label{eq:cmt_opt_mode1}
\frac{dA_1}{dz}=j\omega_1\kappa_{12}B^*e^{j\Delta Kz}A_2-\frac{\alpha_\text{opt}}{2}A_1
\end{equation}

\begin{equation}
\label{eq:cmt_opt_mode2}
\frac{dA_2}{dz}=j\omega_2\kappa_{21}Be^{-j\Delta Kz}A_1-\frac{\alpha_\text{opt}}{2}A_2
\end{equation}

\begin{equation}
\label{eq:acst_amp}
B=B_0e^{-\frac{\alpha_\text{acst}}{2}z}
\end{equation}

\noindent where $\kappa_{ij}$ denotes the optomechanical coupling between mode $i$ and mode $j$ induced by the presence of the acoustic wave, $\Delta K=K-(\beta_1-\beta_2)$ is the wave-vector mismatch (phase mismatch) of the interaction, $\alpha_\text{opt}$ is the optical propagation loss rate, and $\alpha_\text{acst}$ is the acoustic propagation loss rate. 

We have here implicitly assumed that the acoustic wave is undepleted, thereby avoiding $A_i$ terms in Eq.~\eqref{eq:acst_amp}. Because Brillouin scattering is a particle conserving process \cite{boyd2003nonlinear}, if one desired to convert 1$\,$mW of optical power between two optical modes with a 10$\,$GHz acoustic wave, only $\frac{\Omega}{\omega}P_\text{opt}=5\,$nW of acoustic power is required. Thus, for an acoustic drive power more than $1\,${\textmu}W, this assumption is justified.

\subsection{Optomechanical Coupling}
\label{subsec:om_coup}

The optomechanical coupling $\kappa_{ij}$ is mainly induced by two effects: the photoelastic effect through which the strain induced by the acoustic wave directly modifies the permittivity within the optical mode fields, and the moving boundary effect through which the acoustic mode displaces the waveguide boundaries and effectively changes the permittivity seen by the optical modes at those boundaries. The coupling coefficient can therefore be separated into its components as $\kappa_{ij}=\kappa^{PE}_{ij}+\kappa^{MB}_{ij}$. Additionally, through symmetry properties it can be seen that $\kappa_{ij}=\kappa_{ji}^*$ \cite{wolff2015stimulated}. Choosing to define $\kappa = \kappa_{12}$, the photoelastic coupling contribution can be written as

\begin{equation}
\label{eq:kappa_pe}
\kappa^{PE}=\int \epsilon_0\epsilon_r^2\sum_{ijK}\vec{\mathcal{E}}_{1,i}^*\vec{\mathcal{E}}_{2,j}\dbar{p}_{ijK}\dbar{\mathcal{S}}_KdA
\end{equation}

\noindent where $\epsilon_r$ is the relative permittivity, $\dbar{p}_{ijK}$ is the photoelastic tensor in partially reduced notation, and $\dbar{\mathcal{S}}$ is the strain corresponding to the mode shape $\vec{\mathcal{U}}$ in reduced notation ($\dbar{\mathcal{S}}=\nabla_S\vec{\mathcal{U}}$). This integral is taken over the entire cross-section in which the optical and acoustic modes have non-negligible amplitude.

The moving boundary coupling contribution can be written as

\begin{equation}
\label{eq:kappa_mb}
\kappa^{MB}=\int \left(\vec{\mathcal{U}}^*\cdot\hat{n}\right)\left[\left(\epsilon_a-\epsilon_b\right)\vec{\mathcal{E}}_{1,T}^*\cdot\vec{\mathcal{E}}_{2,T}-\left(\epsilon_b^{-1}-\epsilon_a^{-1}\right)\vec{\mathcal{D}}_{1,n}^*\vec{\mathcal{D}}_{2,n}\right]dl
\end{equation}

\noindent where $\epsilon_a$ is the permittivity inside the boundary and $\epsilon_b$ is the permittivity outside the boundary (thus the boundary-normal vector $\hat{n}$ points from $a$ to $b$). Here, the $T$ subscript denotes the field components parallel to the boundary and $n$ denotes the field component normal to the boundary. $\vec{\mathcal{D}}$ denotes the electric displacement field corresponding to the mode shape $\vec{\mathcal{E}}$ ($\vec{\mathcal{D}}=\epsilon\vec{\mathcal{E}}$). This integral is taken over all contours which define a boundary between two materials.

\subsection{Mode Conversion with Perfect Phase-Matching}

In the ideal case where there is no loss and perfect phase-matching ($\alpha_\text{opt}=\alpha_\text{acst}=\Delta K=0$) the coupled optical mode equations [Eqs.~\eqref{eq:cmt_opt_mode1},\eqref{eq:cmt_opt_mode2}] can be solved to arrive at the expected sinusoidal power-exchange oscillations between the two modes, facilitated by the acoustic wave. For some incident power in mode 1 $P_{in}$ and no incident power in mode 2, these equations are

\begin{equation}
\label{eq:mode1_osc_simple}
P_1 = P_{in}\cos^2\left(\frac{\pi z}{2l^0_c}\right)
\end{equation}

\begin{equation}
\label{eq:mode2_osc_simple}
P_2 = P_{in}\frac{\omega_2}{\omega_1}\sin^2\left(\frac{\pi z}{2l^0_c}\right).
\end{equation}

\noindent Here we have introduced the lossless coupling length $l^0_c$, defined as the distance at which all the light has been converted from the first mode into the second for the case of no acoustic loss

\begin{equation}
\label{eq:coup_leng_long}
l^0_c=\frac{\pi}{2|\kappa||B_0|\sqrt{\omega_1\omega_2}}.
\end{equation}

\noindent Defining an effective optical coupling $\bar{\kappa}\equiv|\kappa||B_0|\sqrt{\omega_1\omega_2}$ analogous to the coupling coefficient used in standard optical waveguide coupled mode theory \cite{haus1984waves}, this relation is simply

\begin{equation}
\label{eq:coup_leng_simple}
l^0_c=\frac{\pi}{2\bar{\kappa}}.
\end{equation}

Here it is worth noting that the effective coupling coefficient contains the acoustic amplitude, pointing to the ability to tune the strength and phase of the effective optical coupling \textit{in situ} via the acoustic power for continuous optimization of isolator performance. In the case of non-negligible acoustic loss, decaying acoustic wave results in a decaying coupling between the two optical modes which could be compensated by re-injecting acoustic power periodically along the waveguide in a phase-coherent manner.

\subsection{Mode Conversion with Imperfect Phase-Matching}

It is now worthwhile to consider the more general condition of imperfect phase-matching. Perfect phase-matching ($K=\beta_1-\beta_2$, $\Omega=\omega_1-\omega_2$) in the backward direction provides mode conversion, and imperfect phase-matching ($K\not=\beta_1-\beta_2$, $\Omega\not=\omega_1-\omega_2$) in the forward direction suppresses it, providing the desired non-reciprocity. For assumed steady state operation, where an acoustic wave with a single tone is applied, the second mode is automatically generated at the phase-matched frequency ($\Omega = \omega_1-\omega_2$) and therefore phase-matching (or lack thereof) can only occur with regards to the propagation constant.

The effects of imperfect phase-matching can be evaluated by solving Eq.~\eqref{eq:cmt_opt_mode1}-\eqref{eq:acst_amp} for the output powers in the absence of loss ($\alpha_\text{opt}=\alpha_\text{acst}=0$) as

\begin{equation}
\label{eq:cms_dk_mode1}
P_1=P_{in}\frac{\cos^2\left(\frac{\pi z}{2l^\prime_c}\right)+\gamma^2}{1+\gamma^2}
\end{equation}

\begin{equation}
\label{eq:cms_dk_mode2}
P_2=P_{in}\frac{\sin^2\left(\frac{\pi z}{2l^\prime_c}\right)}{1+\gamma^2}
\end{equation}

\noindent where we have introduced the modified coupling length $l^\prime_c$ and a factor $\gamma$ which limits the amount of power than can be converted. The modified coupling length can be written as

\begin{equation}
\label{eq:lcmod}
l^\prime_c=\frac{\pi}{2\sqrt{\bar{\kappa}^2+\frac{\Delta K^2}{4}}}
\end{equation}

\noindent and the $\gamma$ factor as

\begin{equation}
\label{eq:gamma}
\gamma = \frac{\Delta K}{2\bar{\kappa}}.
\end{equation}

\subsection{Device Length}
\label{subsec:device_length}

These two equations, Eqs.~\eqref{eq:lcmod}-\eqref{eq:gamma}, show that two main effects manifest for imperfect phase-matching when the phase-mismatch $\Delta K$ becomes comparable to the coupling $\bar{\kappa}$: the power conversion 1) occurs over a shorter period and 2) is limited to lower maximum conversion efficiencies. For the proposed device, the achievable phase-mismatch in the forward direction (given perfect phase-matching in the backward direction) sets the device length, and thus we desire a large $\Delta K$ for a given $\bar{\kappa}$ to minimize the device length. In particular, in the forward (imperfectly phase-matched) direction all power must remain in the first mode ($P_1=P_{in}$, $P_2=0$) whereas in the backward (phase-matched) direction all power must be converted to the second mode ($P_2=P_{in}$, $P_1=0$).

The difference in coupling length can be used to achieve full isolation by choosing the device length $L$ to correspond to a null of the sinusoidal conversion in the imperfectly phase-matched direction ($L=2l_c^\prime$) and a peak of the conversion in the perfectly phase-matched direction ($L=l_c^o$). Assuming the bandwidth-maximizing scenario of identical optical group indices for the two optical modes (described in Sec.~\ref{subsec:dispersion}), the phase-mismatch is a function only of the acoustic frequency and optical group index, $\Delta K=2\Omega n_g/c$. Using this relation we find the device length required for full isolation as

\begin{equation}
\label{eq:dev_length}
L=\frac{\pi\sqrt{3}}{2}\frac{1}{\Omega}\frac{c}{n_g}.
\end{equation}

\noindent A shorter device requires increasing the acoustic frequency or decreasing the optical group velocity (e.g. slow light).

For fabricated devices, where the actual length will differ slightly from the designed device length, the ability to tune the acoustic power \textit{in situ} enables high isolation at the cost of insertion loss. For example, a 1\% decrease in device length can be compensated by an approximately $\sqrt{1\%}$ increase to drive power to match $l^0_c$ to $L_\text{fab}$ [see Eq.~\eqref{eq:coup_leng_long}] to ensure all light is converted in the backward direction.

\section{Triply-Guiding Waveguide Design}
\label{sec:design}

In this section we describe the main design considerations for the active section and present an example cross-section which provides both strong optomechanical coupling for power-efficient operation and is partially optimized for optical bandwidth. For simplicity, we choose a standard silicon photonic cross-section, 220 nm thick silicon.

\subsection{Choice of Optical and Acoustic Modes}
\label{subsec:mode_choice}

The foremost consideration in design of the active section is choice of the optical and acoustic modes, and corresponding dispersions, such that the modes are both coupled efficiently and with perfect phase-matching in the backward direction (as described in Sec.~\ref{sec:math}). In particular, this requires that the modal symmetries allow for non-zero optomechanical coupling ($\kappa\not=0$).

As the waveguide width is increased the number of guided optical modes also increases, raising the potential for modal cross-talk which could degrade isolator's performance. It is therefore desired to minimize the number of optical modes, leading to a natural choice of the lowest order, lateral TE modes: TE$_{0,0}$, TE$_{1,0}$, TE$_{2,0}$. Henceforth we drop the second (vertical) mode index and refer to these modes as TE0, TE1, TE2.

As discussed in Sec.~\ref{sec:design}, higher order acoustic modes have cutoff frequencies \cite{Auld1990}, near which they have very low group velocities and are effectively non-propagating. For acoustic wavelengths comparable to the optical wavelength and waveguide widths narrow enough to support only a few optical modes, nearly all higher order acoustic modes will be near cutoff and therefore difficult to use for the propagating architecture desired. This will motivate the usage of one of the three lowest order acoustic modes (polarizations): pressure/longitudinal, horizontal shear, and vertical shear \cite{Auld1990,dostart2017acoustic}.

Out of the three lowest order acoustic modes, horizontal shear can couple TE0/TE1 but not TE0/TE2 whereas the reverse is true for pressure and vertical shear waves. Here, we will choose a horizontal shear wave and TE0/TE1 for our example cross-section to avoid any extraneous optical modes and simplify the analysis.

\subsection{Co-confinement of Optical and Acoustic Modes}
\label{subsec:coconfinement}

The triply-guiding waveguide must also co-confine the optical and acoustic waves so as to maximize the optomechanical coupling. Notably, the basic silicon photonic material set (Si and SiO$_2$) superficially forbids evanescent co-confinement of optical and acoustic waves. Silicon has a higher optical refractive index than its oxide, but the inverse is true in the acoustic domain. Several approaches have been demonstrated to enable co-confinement of optical and acoustic waves including `fin' waveguides \cite{sarabalis2017release}, combined photonic/phononic crystals \cite{balram2016coherent}, or ridge waveguides in a mutually high index material \cite{poulton2012design}. 

A simple and time-tested method for co-confining optical and acoustic waves, and by far the most common, is a suspended waveguide (or membrane) \cite{van2015interaction,van2018electrical,li2015nanophotonic,sohn2018time}. Suspended waveguides (beams) enable relatively larger sidewall displacements and correspondingly higher optomechanical coupling, at the cost of extra fabrication steps. We use a suspended silicon beam here to minimize the acoustic power and demonstrate the potential for high energy-efficiency with acoustics-based optical isolators.

\begin{figure}[t]
	\centering
	\includegraphics[width=.8\columnwidth]{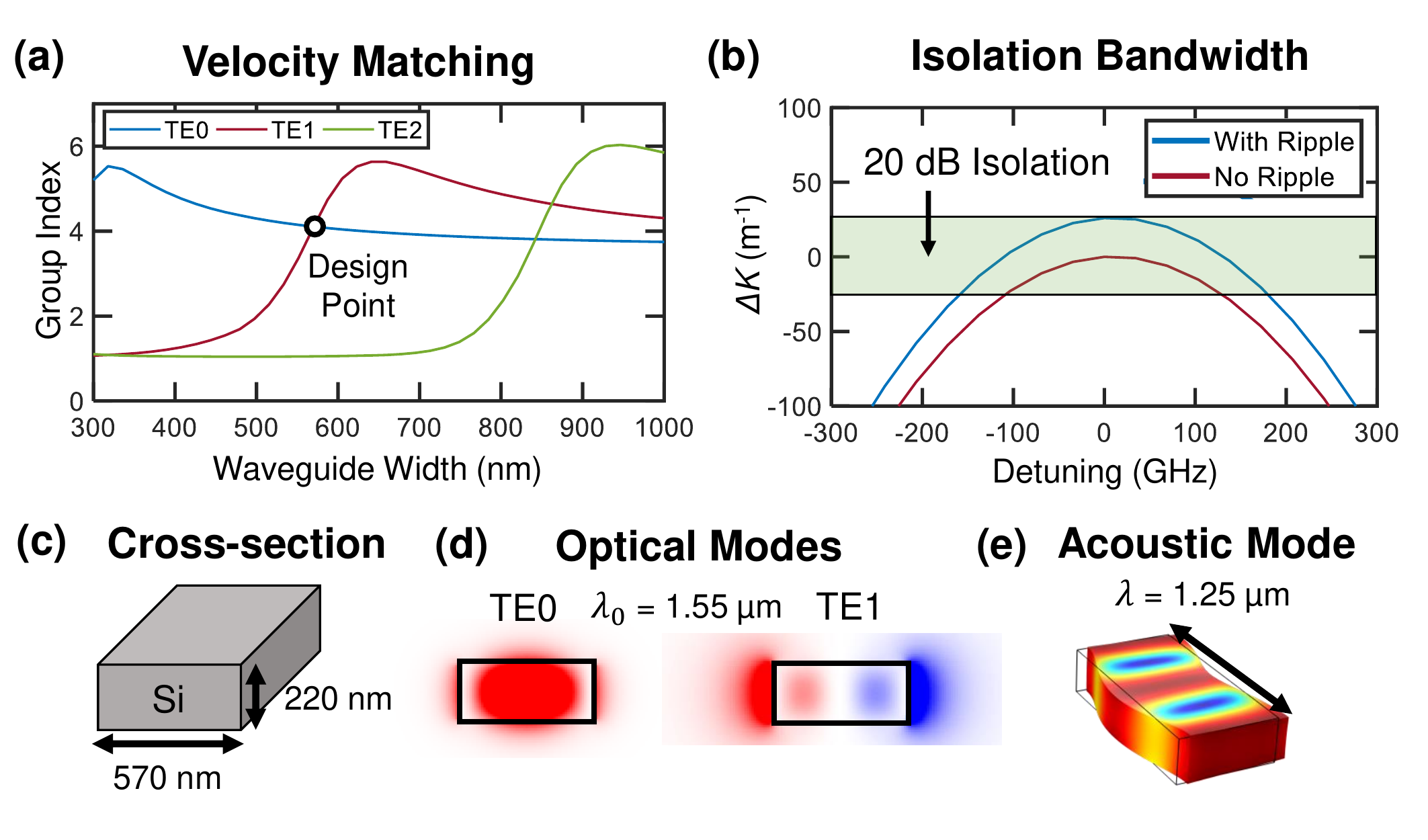}
	\caption{(a) Group index of the first three TE modes; dispersion engineering using the beam width ensures the two optical modes have identical group velocities. (b) Group velocity dispersion creates a quadratically increasing phase-mismatch $\Delta K$ which can be offset by detuning the acoustic wave to create a `ripple' in the isolation which increases the bandwidth. (c) Active region cross-section at the velocity-matched design point. (d) Optical modes of the chosen cross-section at 1550 nm. (e) The phase-matched acoustic mode.}
	\label{fig:dev_design}
\end{figure}

\subsection{Dispersion Engineering: Maximizing Optical Bandwidth}
\label{subsec:dispersion}

For a specific cross-section, and associated optical and acoustic modes, the acoustic excitation wavelength must be chosen to facilitate perfect phase-matching in the backward direction. At some center wavelength (here $\lambda_0=1550$ nm) the effective indices of the two optical modes $n_\text{eff,1}$, $n_\text{eff,2}$ can be found and the phase-matched acoustic wave has an effective wavelength of approximately $2\pi/K=\lambda_0/(n_\text{eff,1}-n_\text{eff,2})$.

Maximizing optical bandwidth requires matching the group velocity of the two optical modes so that the effective indices are independent of frequency (to first order) and phase-matching is maintained over a larger bandwidth [shown for a 220nm thick silicon beam in Fig.~\ref{fig:dev_design}(a)]. Bandwidth will then be limited by the group velocity dispersion, which will cause a quadratic dependence of phase-mismatch on detuning. If the quadratic curve is shifted (`bias detuned' by excitation with an imperfectly phase-matched acoustic wave) in the opposite direction of its curvature, then a wider range of optical frequencies will have low enough phase-mismatch to provide some minimum isolation, creating a ripple in the isolation. The quadratic dependence of phase-mismatch on detuning from center frequency, and the effect of bias detuning, is shown in Fig.~\ref{fig:dev_design}(b).

For the example 220 nm silicon suspended beam we find the optical modes using a standard optical mode solver in Lumerical, while the acoustic mode is found using an equivalent acoustic mode solver \cite{dostart2017acoustic}. For the chosen mode triplet the velocity-matched beam width is 570 nm [Fig.~\ref{fig:dev_design}(c)], corresponding to a $n_g=4.11$ group index. The optical modes at 1550 nm are shown in Fig.~\ref{fig:dev_design}(d), which have effective indices of $n_\text{eff,1}=2.47$, $n_\text{eff,2}=1.23$. The phase-matched acoustic wave for this configuration [Fig.~\ref{fig:dev_design}(e)] has frequency $\Omega/(2\pi)=3.12$ GHz and effective wavelength $2\pi/K=1.25$ {\textmu}m. Using the methods described in Sec.~\ref{subsec:om_coup}, we find the optomechanical coupling coefficient as $\kappa=2.66\times10^{-10}$ m${-1}$sW\textsuperscript{-1/2}. Using Eq.~\eqref{eq:dev_length} the active section length can be found as $L=1.01$ cm. It is worth noting here that any isolator based on mode conversion, such as the isolator proposed here, requires a similarly long device length for broadband operation and GHz driving frequencies (see e.g. \cite{yu2009complete,lira2012electrically,poulton2012design}). We numerically solve for the bias detuning $\delta K$ at which the isolation ratio is 20 dB as $\delta K=26$ rad/m [delimited by the green region in Fig.~\ref{fig:dev_design}(b)], corresponding to a $\sim10$ kHz detuning of the acoustic drive signal from the phase-matched frequency.

\begin{figure}[t]
	\centering
	\includegraphics[width=.7\columnwidth]{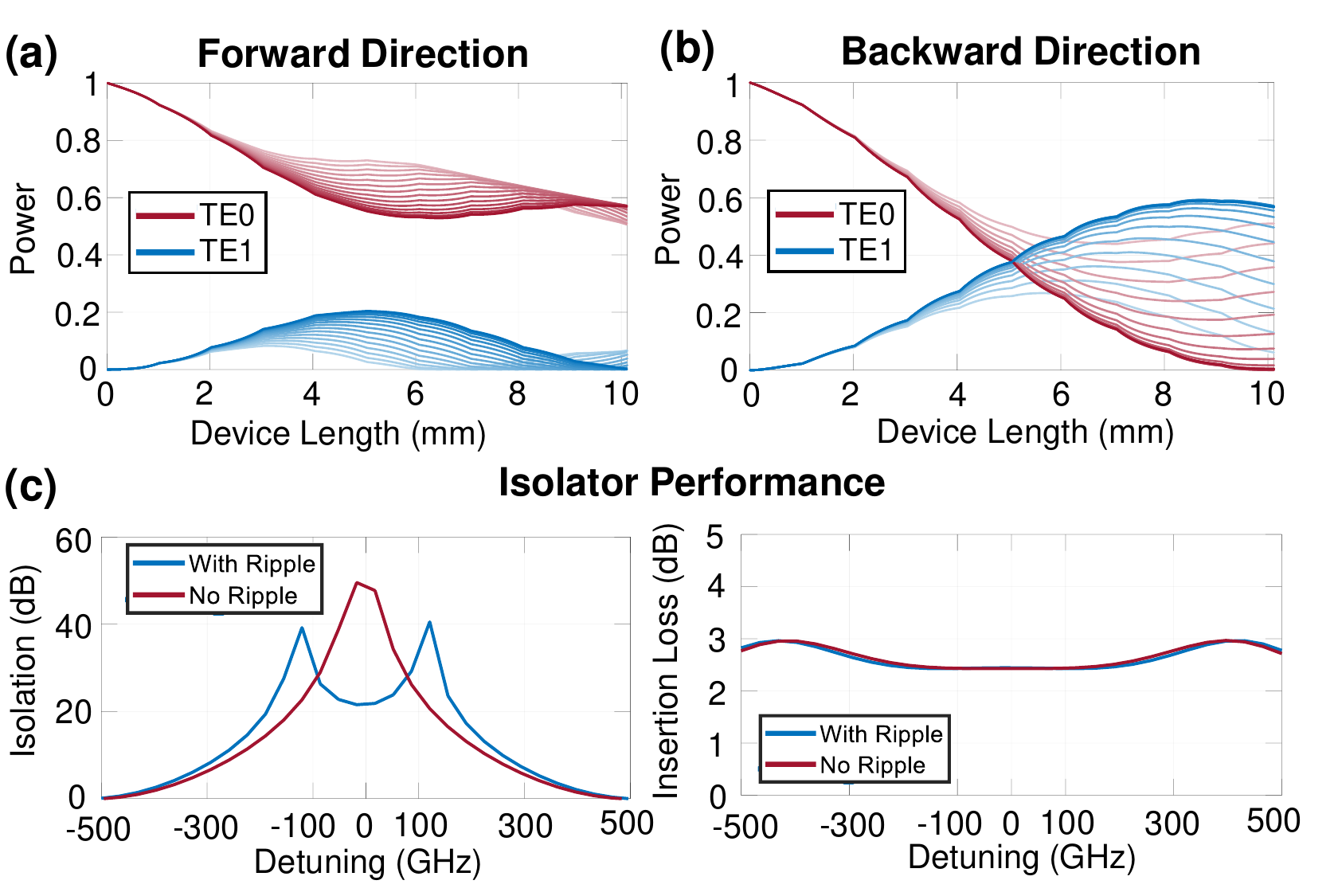}
	\caption{Simulated modal power conversion in (a) forward and (b) backward directions including loss and dispersion. Detuning from center wavelength is shown in lighter shades in 30 GHz increments up to 480 GHz. (c) Device isolation (left) and insertion loss (right), with and without the bandwidth-maximizing ripple.}
	\label{fig:iso_perf}
\end{figure}

\section{Simulated Performance}
\label{sec:simulation}

We evaluate the device's performance for the design presented in Sec.~\ref{sec:design} by numerically solving Eqs.~\eqref{eq:cmt_opt_mode1}-\eqref{eq:acst_amp} along the propagation direction with a 1D finite-difference approach. We assume some device parameters based on previously demonstrated values: an acoustic loss rate of $\alpha_\text{acst}=1$ mm\textsuperscript{-1} (87 dB/cm) \cite{li2015nanophotonic}, an optical loss rate of $\alpha_\text{opt}=2.4$ dB/cm \cite{bogaerts2005nanophotonic}, and 0.07 dB insertion loss per taper \cite{dostart2020serpentine}.

We assume a piezoelectric transducer with efficiency of -14 dB \cite{sohn2018time} since low-loss piezoelectrics have been co-integrated with CMOS photonics \cite{eltes2016low}. However, use of non-piezoelectric transducers, previously demonstrated in both silicon photonic \cite{van2018electrical} and even CMOS \cite{marathe2014resonant} platforms, could enable entirely monolithic (and `zero-change' \cite{sun2015single}) CMOS photonic isolators, and would be interesting to consider in follow-on work.

The acoustic loss rate of 1 mm\textsuperscript{-1} limits interaction length to much less than the 1 cm required by this configuration. We add a new transducer every 1 mm, which in combination with the mechanical limitations of suspended beams leads us to split the 1 cm active length into ten 1 mm suspended beam sections. Each 1 mm active section is preceded by a 30 {\textmu}m long AO mux to re-inject the acoustic wave and anchor the beam. To achieve phase-matching each transducer must be driven with a controlled phase, which could be achieved with on-chip feedback control such as that demonstrated in \cite{sun201645} for thermal control of ring resonances.

The AO mux \cite{muxarxiv} has a 500 MHz acoustic bandwidth within which acoustic insertion loss varies between 0 and 1 dB and forwards/backwards acoustic injection asymmetry is at least 30 dB. It is also optically transparent, with reflection and insertion loss of TE0 less than -40 dB and 0.01 dB respectively, and modal cross-talk less than -70 dB. An additional acoustic insertion loss of about 3 dB is expected between the transducers and the AO mux.

\subsection{Isolator Performance}
\label{subsec:iso_perf}

We numerically simulate mode conversion in both the forward [Fig.~\ref{fig:iso_perf}(a)] and backward [Fig.~\ref{fig:iso_perf}(b)] directions, accounting for loss and dispersion (material and geometrical). For this implementation we predict an insertion loss of 2.6 dB, an isolation of at least 20 dB over a 370 GHz (280 GHz without ripple, a 30\% improvement) bandwidth [Fig.~\ref{fig:iso_perf}(c)], and 1.14 mW of projected electrical drive power. Without the bandwidth-maximizing ripple, at least 30 dB of isolation is predicted at which point mode reflection by the AO mux would begin to limit performance. In theory, perfect isolation is obtainable with the case of a perfect AO mux, and our simulations predict up to 50 dB of isolation.

\begin{figure}[t]
	\centering
	\includegraphics[width=\columnwidth]{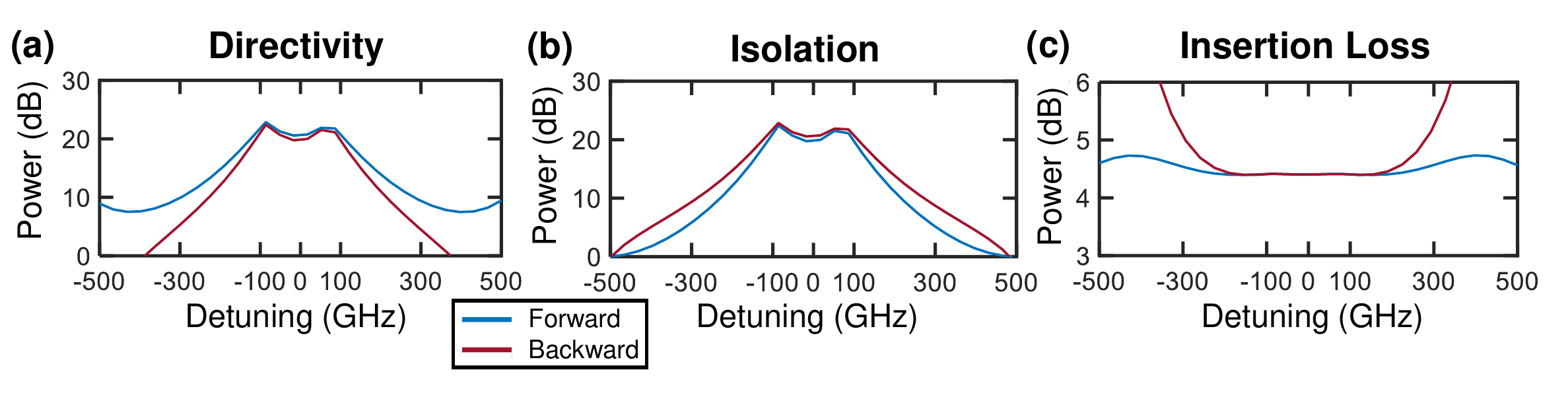}
	\caption{\textbf{a} Circulator directivity in both directions, denoting the ratio of power entering the correct output mode relative to the incorrect output mode (e.g. $|S_{31}|^2/|S_{41}|^2$). \textbf{b} Circulator isolation in both directions (forward -- $|S_{31}|^2/|S_{13}|^2$, $|S_{42}|^2/|S_{24}|^2$; backward -- $|S_{23}|^2/|S_{32}|^2$, $|S_{14}|^2/|S_{41}|^2$). \textbf{c} Circulator insertion loss in both directions between the desired ports. All plots include bandwidth-maximizing ripple and effects of two mode multiplexers in forward and backward directions.}
	\label{fig:circ_perf}
\end{figure}

\subsection{Circulator Performance}
\label{subsec:circ_perf}

The EMP isolator can be extended to a 4-port circulator by replacing the tapers with optical multiplexers (resulting in one at either end of the non-reciprocal section, see Fig.~\ref{fig:iso_concept}(b)). For the optical multiplexers we assume -30 dB crosstalk and $\sim1$ dB of insertion loss over a $\sim2$ THz bandwidth \cite{sun2016ultra}. The circulator requires no more power than the isolator, but the performance is slightly degraded by modal cross-talk and insertion loss in the multiplexers. In order to maintain 20 dB isolation the circulator requires a smaller bias detuning of $\delta K=10$ rad/m, resulting in a smaller bandwidth. The simulated circulator performance is shown in Fig.~\ref{fig:circ_perf}, where isolation and directivity of at least 20 dB are expected over a 220 GHz bandwidth with 4.5 dB optical insertion loss. In Fig.~\ref{fig:circ_perf} we plot directivity, isolation, and insertion loss with regards to the forward and backward directions, as both modes propagating in a given direction experience the same directivity/isolation/insertion loss. For the circulator, isolation is defined as the ratio of transmissions to two output ports, one in the desired direction and one in the undesired direction (for example $|S_{31}|^2/|S_{13}|^2$, for port labels see Fig.~\ref{fig:iso_concept}). Directivity is defined as the ratio of transmission from a given input port to the desired output port and transmission between the same input port and an undesired output port (for example $|S_{31}|^2/|S_{41}|^2$). Because we assume negligible reflections, we do not plot directivities which would require reflection (such as $|S_{31}|^2/|S_{21}|^2$).

\section{Discussion}
\label{sec:discussion}

\begin{table}[t]
	\centering
	\setlength\tabcolsep{5pt}
	\begin{tabular}{cccccccccc}
		\hline
		Parameter & \cite{kittlaus2018nonreciprocal} & \cite{sohn2018time} & \cite{tzuang2014non} & \cite{lira2012electrically} & \cite{doerr2014silicon} & \cite{huang2016electrically} & \cite{huang2017integrated} & This Work\\
		\hline
		Approach & SBS & EM & OE & OE & OE & MO & MO & EM \\
		Platform & SiP & AlN & SiP & SiP & SiP & Si:CIG & Si:CIG & SiP \\
		Realization & Expt. & Expt. & Expt. & Expt. & Expt. & Expt. & Expt. & Sim. \\
		Broadband & Yes & No & Yes & Yes & Yes & No & Yes & Yes\\
		Length (mm) & >10 & 0.17 & 8.35 & 18.7 & 9 & 0.07 & 1.5 & 10 \\
		Power (mW) & 50 & 100 & 500 & 25 & 2000 & 9.6 & 260 & 1.14 \\
		IL (dB) & 20 & 7.7 & -- & 70 & 4 & 2.3 & 8 & 2.6 \\
		Isolation (dB) & 25 & 15 & 2.4 & 3 & 3 & 32 & 29 & 20 \\
		Bandwidth (GHz) & 150 & 1 & $\sim$2,500 & 200 & 4,500 & 11 & 2,300 & 370 \\
		\hline
		\textbf{Energy (fJ/bit)} & 1,300 & 400,000 & 800 & 500 & 1,800 & 3,500 & 450 & \textbf{12.0} \\
		\hline
	\end{tabular}
	\caption{Comparison to other representative integrated photonic isolator demonstrations. We abbreviate the approach to creating a time-varying permittivity as: SBS -- stimulated Brillouin scattering, EM -- electro-mechanical transducers, OE -- optoelectronic modulators, MO -- magneto-optic. The material platform used for the demonstration is abbreviated as: SiP -- silicon photonic, AlN -- aluminum nitride, Si:CIG -- hybrid silicon-on-insulator with wafer-bonded cerium substituted yttrium iron garnet.}
	\label{tb:comp}
\end{table}

For comparison to other on-chip (no external magnet) isolation approaches, we have collected some representative devices in Table~\ref{tb:comp}. To estimate energy per bit efficiencies, we have assumed 0.25 bps/Hz spectral utilization. Further comparison of non-resonant approaches to on-chip isolation can also be found in \cite{yang2014experimental}. The proposed EMP isolators offer a self-contained photonic component with equivalent function to a passive isolator, and no magneto-optic materials. This comes at the cost of non-zero electrical power consumption, but as shown this power is expected to be extremely low relative to other relevant components. The EMP isolator only requires about 1 mW of electrical drive power, whereas the best demonstrated approach so far (magneto-optic) requires nearly 10 mW to power the electro-magnet to induce non-reciprocity. Energy per bit is also a useful metric for data transmission links in intra- and inter-chip optical communication applications \cite{miller2017attojoule}. The proposed isolator example is also more than an order of magnitude more efficient in terms of energy per bit than the next-best on-chip approach, either resonant or non-resonant. Being in the 10 fJ/bit regime in energy cost, it adds a negligible footprint to the energy budget of integrated photonic communication links, where modulators and detectors average 10s to 100s of fJ/bit efficiencies in the latest approaches, respectively \cite{sun2015single}. Many previous on-chip isolator demonstrations would dominate the power budget of such a link. Future on-chip systems would therefore benefit from the low power usage of EMP isolators, which is due to the high energy efficiency of electrically-transduced acoustic waves.

Further work is needed to experimentally validate this concept, as well as to investigate improved designs that achieve optical bandwidths above 1 THz, pure CMOS acoustic wave excitation, and shorter device lengths. The bandwidth can be further increased by minimizing group velocity dispersion using more sophisticated cross-section designs such as \cite{poulton2012design}. In general, shorter device lengths will decrease design sensitivity to phase-mismatch and will thereby widen the bandwidth. A sub-mm active section could be achieved by a combination of increasing the drive frequency (e.g. from 3 GHz to 10 GHz) and using slow wave (e.g. periodic photonic bandgap) structures to decrease the optical group velocity, increasing the bandwidth to multiple THz. This, of course, must be accomplished while keeping reflections low. In particular, a device length below the acoustic decay length significantly increases the feasibility of the design as only one suspended section, AO mux, and transducer would be needed. We expect that future designs in silicon photonic platforms will also take advantage of transducers which do not require piezoelectrics, such as \cite{van2018electrical,marathe2014resonant}. Combined, these advances can enable an exceptionally high-performing, compact, and efficient optical isolator using only CMOS-compatible materials.

In this paper we introduce the concept of an EMP isolator, theoretically evaluate its operation, discuss design considerations, and demonstrate the potential for outstanding performance and low power usage through simulations. The example design presented here improves over other on-chip isolator demonstrations through use of co-propagating acoustic and optical waves in a single cross-section for lossless, non-reciprocal mode conversion by the acoustic wave and efficient use of the acoustic power for mode conversion. Simulations indicate that the proposed EMP isolator requires only 1.14 mW of electrical drive power to provide at least 20 dB of isolation across a 370 GHz bandwidth with 2.6 dB of insertion loss. The EMP isolator can also be configured as a 3- or 4-port circulator, where a decreased bandwidth of 220 GHz and increased insertion loss of 4.5 dB result from the necessity of two mode multiplexers. Future work can enable far more compact, easily controlled, and CMOS-friendly implementations, shrinking the 1 cm long design here to below 1 mm, potentially leading to energy-efficient isolators based on electrically-driven acoustic waves becoming standard silicon photonic foundry components.

\section*{Funding Information}

National Science Foundation Graduate Research Fellowship Grant (1144083); Packard Fellowship for Science and Engineering (2012-38222).

\bibliographystyle{ieeetr}
\bibliography{biblio}

\end{document}